\documentclass[twocolumn,showpacs,preprintnumbers,footinbib]{revtex4}
\usepackage{bm,enumerate,dcolumn,tikz,graphicx,color,amsmath,amssymb}
\usepackage{epstopdf}
\usepackage{empheq}
\usepackage{capt-of}
\usepackage{pgfplots}

\usepgfplotslibrary{colormaps,external}
\usetikzlibrary{decorations.markings}
\usetikzlibrary{arrows}

\newcommand{\comment}[1]{}
\newcommand{\bra}{\langle}
\newcommand{\ket}{\rangle}

\newcommand{\velg}{{\textsc{\tiny VG}}}

\definecolor{lightblue}{rgb}{.8, .8, 1}
\definecolor{myyellow}{RGB}{254,241,24}
\definecolor{myorange}{RGB}{234,125,1}
 
\usepackage{array}
\newcolumntype{C}[1]{>{\centering\let\newline\\\arraybackslash\hspace{0pt}}m{#1}}
\newcolumntype{L}[1]{>{\raggedright\let\newline\\\arraybackslash\hspace{0pt}}m{#1}}
\newcolumntype{R}[1]{>{\raggedleft\let\newline\\\arraybackslash\hspace{0pt}}m{#1}}

\newcommand{\threej}[6]{
\begin{pmatrix}
  #1 & #2 & #3\\
  #4 & #5 & #6
\end{pmatrix}
}

\begin{document}

\title{Tune-out wavelengths and landscape modulated polarizabilities of alkali Rydberg 
atoms in infrared optical lattices}  
\author{Turker Topcu and Andrei Derevianko}
\affiliation{Department of Physics, University of Nevada, Reno, NV 89557, USA}
\date{\today}

\begin{abstract}
Intensity modulated optical lattice potentials can change sign for an alkali metal 
Rydberg atom, and the atoms are not always attracted to intensity minima in optical lattices with 
wavelengths near the CO$_2$ laser band. Here we demonstrate that such IR lattices can be tuned so 
that the trapping potential seen by the Rydberg atom can be made to vanish for atoms in ``targeted" 
Rydberg states. Such state selective trapping of Rydberg atoms can be useful in controlled cold 
Rydberg collisions, cooling Rydberg states, and species-selective trapping and transport of Rydberg 
atoms in optical lattices. 
We tabulate wavelengths at which the trapping potential vanishes for the $n$s, $n$p, and $n$d 
Rydberg states of Na and Rb atoms, and discuss advantages of using such optical lattices for state 
selective trapping of Rydberg atoms. We also develop exact analytic expressions for the lattice 
induced polarizability for the $m_z=0$ Rydberg states, and derive an accurate formula predicting 
tune-out wavelengths at which the optical trapping potential becomes invisible to Rydberg atoms in 
targeted $l=0$ states. 
\end{abstract}


\maketitle

\section{Introduction}

Recent advances in quantum computing with Rydberg atoms, and quantum simulation experiments 
studying many-body systems have been possible thanks to developments in optical trapping 
methods~\cite{SafWalMol10,WeiMulLes10,BloDalNas12}. Many of these experiments involve cold 
Rydberg atoms, such as implementations of quantum logical gates to realize scalable quantum 
computing~\cite{SafWalMol10,jaksch}, and rely on optically 
trapped cold Rydberg atoms. Because of this, optical traps have been studied  
\cite{dutta,saffman,youngeNJP,anderson,zhang} as a method of trapping Rydberg atoms, alongside 
other trapping schemes involving static electric~\cite{mosley} and magnetic fields~\cite{mayle}. 
The main advantage of optical trapping stems from small dynamic Stark shifts experienced 
by the atomic states (MHz) compared to the shifts in electrostatic traps (GHz)~\cite{SafWalMol10}. 

Among various methods in optical trapping toolbox is the ``tune-out" phenomena, where the optical 
lattice frequency is tuned to make it invisible to a once trapped atom. This method was first 
proposed by LeBlanc {\it et al.}~\cite{LeBThy07}, where an optical lattice is tuned to 
a wavelength at which the lattice potential vanishes for a given atomic species in ground state. 
This enables one to design species-specific optical lattices, where one atomic species is trapped 
and the other is ``tuned out". Such selectivity can be also achieved for different low-lying levels 
of the same atom, providing a state-specific optical lattice. These types of optical lattices have 
been investigated especially for ground state alkali metal atoms~\cite{ArpSafCla11}, and several 
applications have emerged over time, such as quantum computing schemes utilizing a storage lattice 
for encoding qubits and a transport lattice for addressing individual qubits~\cite{DalBoyYe08}, 
state-selective transport~\cite{GajOpaDas11}, and cooling in strongly correlated optical 
lattices~\cite{McKDeM11}. 

The optical potential for off-resonant light vanishes for an atom in a specific 
state when its dynamic polarizability vanishes. This makes it straightforward to realize species- 
and state-specific optical lattices for ground state atoms because the resonant structure at 
low-lying levels provides many frequencies at which the atomic polarizability vanishes. On the 
other hand, species-specific lattices have not been realized for Rydberg atoms, because 
the common wisdom is that the polarizability of a Rydberg state is essentially that of a free 
electron, which is always negative (although there has been evidence to the contrary 
\cite{YouKnuAnd10}). Since the free electron polarizability does not change sign, it never 
vanishes, making it impossible to tune-out Rydberg atoms from optical lattices. 
In this paper, we demonstrate that this is not true, and tune-out wavelengths exist at 
which the optical lattice becomes invisible to a Rydberg atom in IR 
lattices with wavelengths up to $\sim$10$^4$ nm. We tabulate a few of these tune-out 
wavelengths for some Rydberg states of Na and Rb atoms, and derive a simple analytical 
expression which accurately predicts half of the available tune-out wavelengths 
for alkali-metal atoms. 

The trapping potential seen by the Rydberg electron is composed of an intensity modulated 
``landscaping" term and a constant offset, which does not depend on the position of the 
atom along the lattice~\cite{TopDerLands}. This landscaping term modulates the free electron 
polarizability, and causes trapping potential to change sign. Because the trapping potential 
has to vanish in order to change sign, this allows us to tune out the lattice for Rydberg atoms 
in one state, while another state remains trapped. 

In the next section, we provide a description of the tune-out phenomena for Rydberg atoms 
trapped in optical lattices. To this end, we use a one-dimensional toy model to introduce 
the underlying mechanism, and then proceed to rigorous treatment and calculation of the tune-out 
wavelengths for Na and Rb atoms. We then turn out attention to three-dimensional optical 
lattices in Sec. III. We evaluate intensity modulated polarizabilities in 3D lattices, and argue 
that tune-out can be achieved for some states in all directions. Finally, we derive 
approximate analytic expressions for intensity modulated polarizabilities and tune-out 
wavelengths, which accurately predict half of the available tune-out wavelengths. 
We use atomic units throughout this paper unless we explicitly state otherwise. 

\section{Tune-out frequencies for Rydberg states}

\begin{figure}[h!tb]
	\begin{center}
		\resizebox{85mm}{!}{\includegraphics{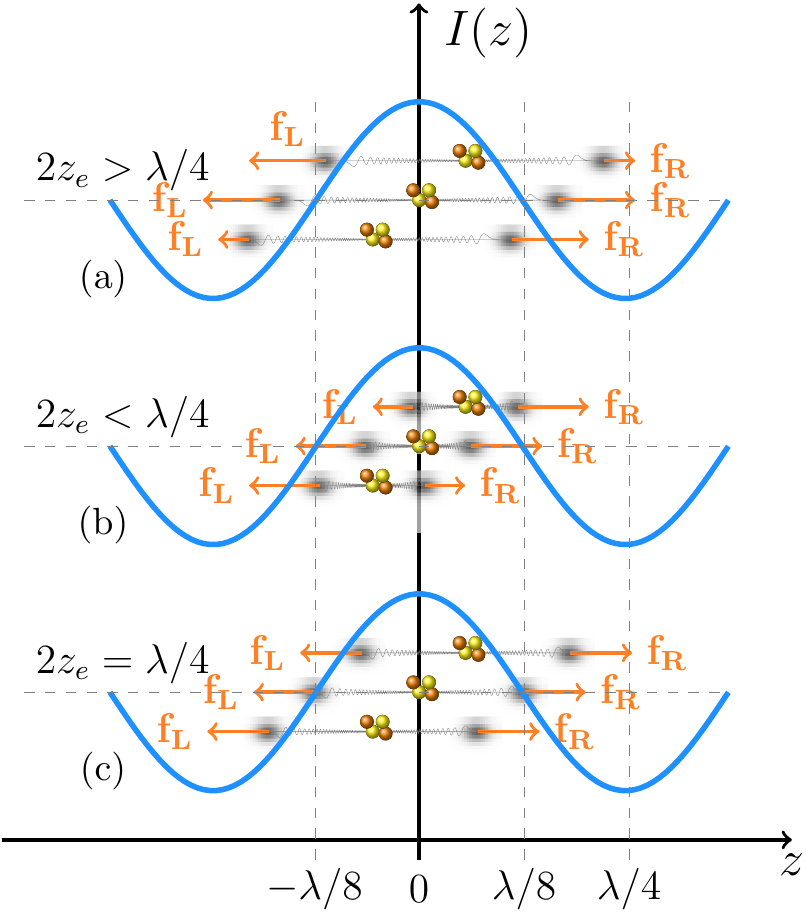}} 
  \end{center}
	\caption{Toy model of a Rydberg atom in one-dimensional lattice, where the 
	vertical axis is the laser intensity $I(z)$. When the size of the 
	Rydberg orbit $2z_e^0=\lambda/4$, 
	the forces $\mathbf{f}_R$ and $\mathbf{f}_L$ acting 
	on the localized "lumps" of electron density are equal in magnitude and 
	the net force acting on the atom is zero: this is the tune-out scenario. 
	}
	\label{fig:opt_lattice}
\end{figure}

We start our description of the tune-out frequencies for Rydberg states using a toy model 
of a Rydberg atom in a one-dimensional optical lattice. In this model, the atom is one-dimensional 
and the electron density is concentrated on either side of the atom near the classical turning 
points at $z_{\rm e}=\pm2n^2$. This toy model of the Rydberg atom in a one-dimensional lattice is 
discussed in detail in~\cite{TopDerLands}, illustrating the interplay between the Rydberg orbit size 
and the lattice wavelength. Two cases when the size of the Rydberg orbit is larger and smaller 
than $\lambda/4$ correspond to cases in which the laser intensity maxima are the stable and unstable 
equilibrium points of the optical potential. As the wavelength of the lattice is varied from 
$\lambda/4 >4n^2$ to $\lambda/4 <4n^2$, there is the critical case when $\lambda/4 =4n^2$ 
(Fig.~\ref{fig:opt_lattice}). In this case, if the atom is displaced from the intensity maxima, one 
side falls inside the nearby inflection surface while the other falls outside. Because of the symmetry 
of the intensity distribution, the optical dipole forces $\mathbf{f}_R$ and $\mathbf{f}_L$ are equal 
in magnitude, and the atom feels no net tug, making it blind to the optical lattice. This is the 
``tune-out" situation we are interested in. 

We will now derive explicit expressions for the dynamic polarizabilities of Rydberg states, 
which will allow us to calculate tune-out wavelengths for alkali-metal Rydberg atoms. 
The electric field of two linearly polarized counter-propagating laser beams 
can be used to construct a one-dimensional optical lattice:  
\begin{equation}\label{field}
\begin{split}
&\mathbf{F}  = 2F_0\hat{\epsilon} \sin(\mathbf{k}\cdot\mathbf{r})\cos(\omega t) \;,
\end{split}
\end{equation}
where $k=\omega/c=2\pi/\lambda$ is the wave vector, $\mathbf{r}$ is the coordinate 
of the electron, $F_0$ is the electric field strength, and $\hat{\epsilon}$ is the polarization 
direction of the laser beams. 
We will work in the velocity gauge (also known as the transverse or the Coulomb gauge). 
In this gauge, the scalar potential is zero, and 
the vector potential $\mathbf{A}_{\velg}$ and the field are related by 
$\mathbf{F}=-\partial\mathbf{A}_{\velg}/\partial t$. 
Then Hamiltonian for the electron is 
\begin{equation}\label{eq:hamiltonian}
H = \frac{p^2}{2} +V_{C} - \frac{\mathbf{A}_{\velg}\cdot\mathbf{p}}{c} 
	+ \frac{(A_{\velg})^2}{2c^2} \;,
\end{equation}
where $V_C$ is the core potential seen by the electron. 
The vector potential that gives rise to~\eqref{field} in this gauge is 
\begin{equation}
\mathbf{A}_{\velg}(\mathbf{r}, t) = -\frac{2F_0}{\omega} \hat{\epsilon}\;
  \sin(\mathbf{k}\cdot\mathbf{r}) \sin(\omega t) \;.
\end{equation}
Ref.~\cite{TopDerGauge} demonstrated that the last term in the Hamiltonian provides the 
dominant contribution to the energy shift for Rydberg states, which is what we will focus 
on. We use the first order perturbation theory for the $(A_{\velg})^2$ term since it is 
already second order in the field strength. Then we can express the lattice potential for a 
Rydberg state $|r\ket$ $U_r=-\alpha_r(\omega)F_0^2/4$, where $\alpha_r(\omega)$ is 
the dynamic polarizability for the Rydberg state $|r\ket = |nlm_z\ket$ :
\begin{equation}\label{final_pol}
\alpha_{nlm_z}(\mathbf{R},\omega)  = -\frac{1}{\omega^2} 
  \bra nlm_z|\sin^2(\mathbf{k}\cdot\mathbf{r})|nlm_z\ket . 
\end{equation} 

We can express the coordinate of the electron $\mathbf{r}$ in terms of the 
center of mass coordinate of the atom $\mathbf{R}$ and the coordinate of 
the Rydberg electron relative to the center of mass $\mathbf{r}_e$, {\it i.e.} 
$\mathbf{r}=\mathbf{r}_e+\mathbf{R}$. 
Since only the $\mathbf{R}$-dependent part of the lattice potential 
$U_r=-\alpha_{r}(\omega)F_0^2 /4$ exerts force on the atom, it is 
advantageous to separate out the $\mathbf{R}$-dependent part of the 
polarizability, and refer to it as the trapping potential. 
%
%
Simple trigonometric manipulation leads to 
\begin{equation}
\begin{split}
\alpha_{nlm_z}(\omega) = -\frac{1}{\omega^2} 
  \Big [ &
    \sin^2(\mathbf{k}\cdot\mathbf{R}) 
	  \bra nlm_z|\cos(2\mathbf{k}\cdot\mathbf{r}_e)|nlm_z\ket \\
    &+ \bra nlm_z|\sin^2(\mathbf{k}\cdot\mathbf{r}_e)|nlm_z\ket 
  \Big ] .
\end{split}
\end{equation} 
At this point, we will assume a one-dimensional linearly polarized optical lattice 
propagating along the $\hat{z}$ direction, so that $\mathbf{k}\cdot\mathbf{r}_e = k z_e$. 
We are also assuming that the energy splitting due to quantum defects between the low angular 
momentum states inside the $n$-manifold is much larger than the Stark coupling induced by 
the lattice field. In a one-dimensional lattice, this guarantees that the Rydberg atom stays 
in its initial state when trapped by the lattice field, since $m$ is a good quantum number. 
In such a 1D lattice polarizability becomes 
\begin{equation}\label{eq:lands_pol_split}
\begin{split}
\alpha_{nlm_z}(\omega) = -\frac{1}{\omega^2} 
  \Big [ &
    \sin^2(kZ) 
	  \bra nlm_z|\cos(2k{z}_e)|nlm_z\ket \\
    &+ \bra nlm_z|\sin^2(k{z}_e)|nlm_z\ket 
  \Big ] \;,
\end{split}
\end{equation} 
where $Z$ is the position of the atom along the $z$-axis. We can now separate the 
optical potential in two pieces: 
\begin{equation}\label{eq:pot_split}
U_r(Z)=U^0_r + U^Z_r \sin^2(kZ) \;,
\end{equation} 
where 
\begin{eqnarray}\label{eq:pot_zdep}
U^Z_r &=& \frac{F_0^2}{4\omega^2} \bra n l m_z|\cos (2k z_e)|n l m_z \ket 
	\equiv- \alpha^{\rm lsc}_{nlm_z}(\omega) \frac{F_0^2}{4} \; , \\
U^0_r &=& \frac{F_0^2}{4\omega^2} \bra n l m_z|\sin^2 (k z_e)|n l m_z \ket \; .
\label{eq:pot_offset}
\end{eqnarray}
The term $U^0_r$ is merely an ``offset", and does not depend on where the atom is along the 
optical lattice. Therefore it exerts no force on the atom. It is the second term in 
\eqref{eq:pot_split} that is relevant in trapping the Rydberg atom. Here we introduced the 
intensity ``landscape-averaged polarizability" $\alpha^{\rm lsc}_{nlm_z}(\omega)$, and unlike 
the free electron polarizability $\alpha_{\rm e}=-1/\omega^2$, it can be both positive or 
negative depending on the wavelength. The fact that $\alpha^{\rm lsc}_{nlm_z}(\omega)$ can 
take both positive and negative values means that it has to vanish at certain wavelengths, 
and for these special $\lambda$ the atom would be unaware of the optical lattice. 
This happens when 
\begin{equation}\label{eq:cos_2kz}
\alpha^{\rm lsc}_{nlm_z}(\omega) = -\frac{1}{\omega^2} 
	  \bra nlm_z|\cos(2k{z}_e)|nlm_z \ket = 0 \;,
\end{equation}
which we refer to as the tune-out condition. The landscaping polarizability 
$\alpha^{\rm lsc}_{nlm_z}$ in this 1D lattice  can be explicitly written by expanding 
$\cos(2kz_e)$ in spherical Bessel functions: 
\begin{eqnarray}\label{cos_expansion}
\begin{split}
\alpha^{\rm lsc}_{nlm_z}(\omega) = -&\frac{(2l+1)}{\omega^2} 
		\sum_{l'=\text{even}} (2l'+1) (-1)^{l'/2-m_z} \\
			&\times \threej{l}{l'}{l}{-m_z}{0}{m_z} 
			\threej{l}{l'}{l}{0}{0}{0} \\
			&\times \int_0^\infty dr_{e} P_{nl}^2(r_e) j_{l'}(2kr_e) \;.
\end{split}
\end{eqnarray}
The Rydberg state landscaping polarizabilities $\alpha^{\rm lsc}_{n,l,m_z}(\omega)$ are 
calculated using non-relativistic states computed in a 12,000 a.u. radial box using 
9000 points on a square-root mesh. We calculate $P_{nl}(r)$ by directly integrating the 
one-electron time-independent Schr\"odinger equation using the well known quantum defect 
potentials for the alkali atoms~\cite{gallagher}. Finally, we plot $\alpha^{\rm lsc}_{n,l,m_z}$ 
to search for wavelengths at which the landscaping polarizability vanishes, and the optical 
lattice becomes invisible for the atom in state $|nlm_z\ket$. 

\begin{figure}[h!tb]
	\begin{center}
 		\resizebox{80mm}{!}{\includegraphics{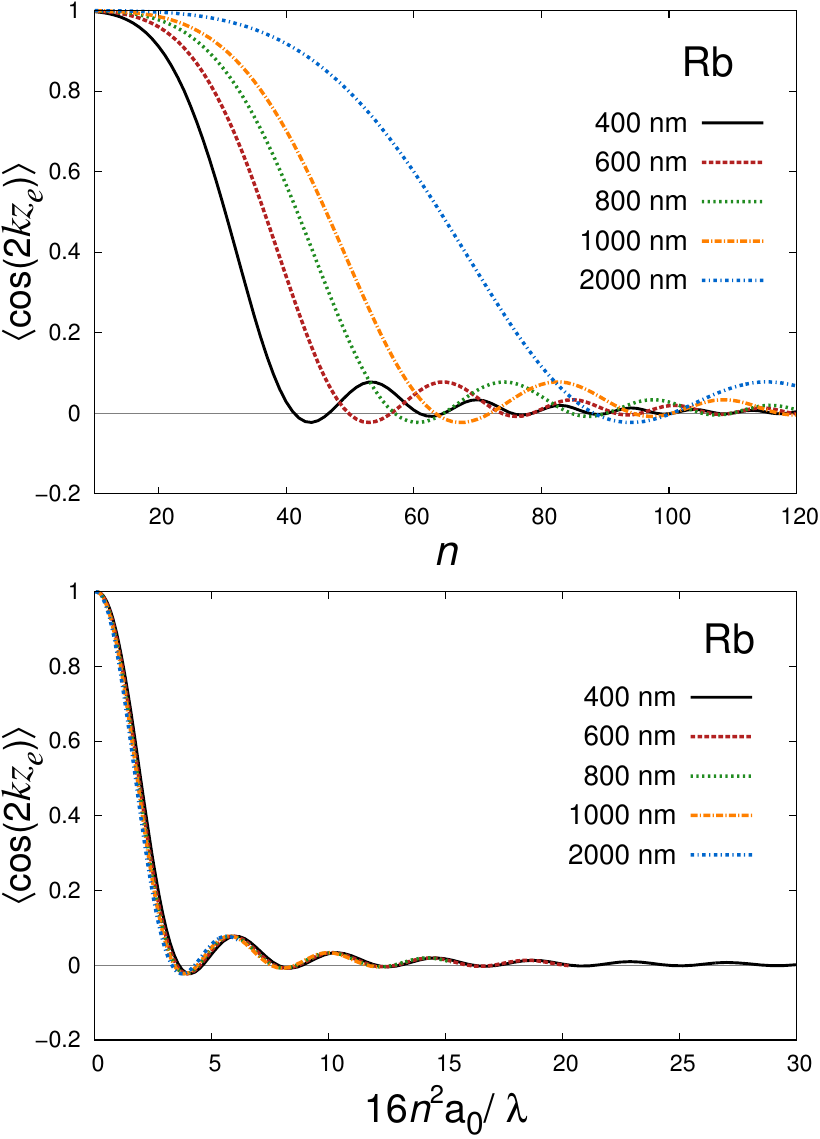}}
  \end{center}
	\caption{The modulation factor $\bra\cos(2kz_e)\ket$ as a function of $n$ for $l=0$
		states of Rb for various $\lambda$. 
		At low $n$, $\bra\cos(2kz_e)\ket\rightarrow 1$ and 
		$\alpha^{\rm lsc}_{nlm_z}(\omega)\rightarrow \alpha_e(\omega)$. At higher $n$,
		$\alpha^{\rm lsc}_r$ oscillates around zero, and each of the wavelengths seen in the 
		figure becomes a tune-out wavelength for an infinite number of $n$ states. 
		The lower panel demonstrates the universal dependence of the factor $\bra\cos(2kz_e)\ket$ 
		from the upper panel on $n^2 a_0/\lambda$. 
	}
	\label{fig:cos_scaled}
\end{figure}

The matrix element $\bra nlm_z|\cos(2k{z}_e)|nlm_z \ket$ in Eq.~\eqref{eq:cos_2kz} 
can be seen as a factor modulating the free electron polarizability $\alpha_{\rm e}=-1/\omega^2$ 
according to the intensity profile of the 1D lattice in the lattice propagation direction: 
$\alpha^{\rm lsc}_{nlm_z}(\omega)=\bra \cos(2kz_e) \ket \alpha_{\rm e}(\omega)$. 
We plot this modulation factor $\bra \cos(2kz_e) \ket$ as a function of $n$ for $l=0$ 
states of Rb atom in Fig.~\ref{fig:cos_scaled} for various wavelengths. In the upper panel, 
all curves start from 1 at small $n$, because in this limit $\bra\cos(2kz_e)\ket\rightarrow 1$ 
and $\alpha^{\rm lsc}_{nlm_z}(\omega) \rightarrow \alpha_{\rm e}(\omega)$. For larger 
$n$, they oscillate with diminishing amplitude towards higher $n$. Although oscillation 
amplitudes become smaller at larger $n$, they still go through $\bra \cos(2kz_e) \ket=0$, 
and each wavelength in the upper panel in Fig.~\ref{fig:cos_scaled} is a tune-out wavelength 
for an infinite number of $n$s-states with $n>40$. Also note that the longer the wavelength the 
higher $n$ it takes to modulate $\alpha_e$.

The oscillations seen in the upper panel of Fig.~\ref{fig:cos_scaled} can be shown to harbor 
a universal character. For $l=0$ states, the modulation factor $\bra\cos(2kz_e)\ket$ from 
Eq.~\eqref{cos_expansion} reduces to 
\begin{eqnarray}\label{j0_expansion}
\bra ns|j_0(2kr_e)|ns\ket &=& \bra ns|\sum_p c_p (kr_e)^p |ns\ket \nonumber \\
	&=& \sum_p c_p k^p \bra ns|r_e^p|ns \ket \;, 
\end{eqnarray}
where we expanded $j_0(2kr_e)$ in power series. In the limit $n\rightarrow \infty$, 
we can use the expression $[\bra ns|r^p|ns\ket]^{1/p} = b_p n^2$, 
where $b_p$ is a coefficient. With this, we can show that 
\begin{eqnarray}\nonumber
\bra j_0(2kr_e)\ket &=& \bra ns|\sum_p c_p b_p (kn^2)^p |ns\ket \\
	&=& F(kn^2) \;. 
\end{eqnarray}
In other words, the landscaping modulation factor $\bra\cos(2kr_e)\ket$ is a function 
of $kn^2$. The bottom panel of Fig.~\ref{fig:cos_scaled} illustrates this behavior 
where all the curves in upper panel are plotted as a function of $16n^2 a_0/\lambda$. 
The fact that all the curves lay on top of each other demonstrates the universal dependence 
on $k\bra r_e\ket \propto n^2 a_0/\lambda$ for all wavelengths. 

\begin{figure}[h!tb]
	\begin{center}
 		\resizebox{85mm}{!}{\includegraphics{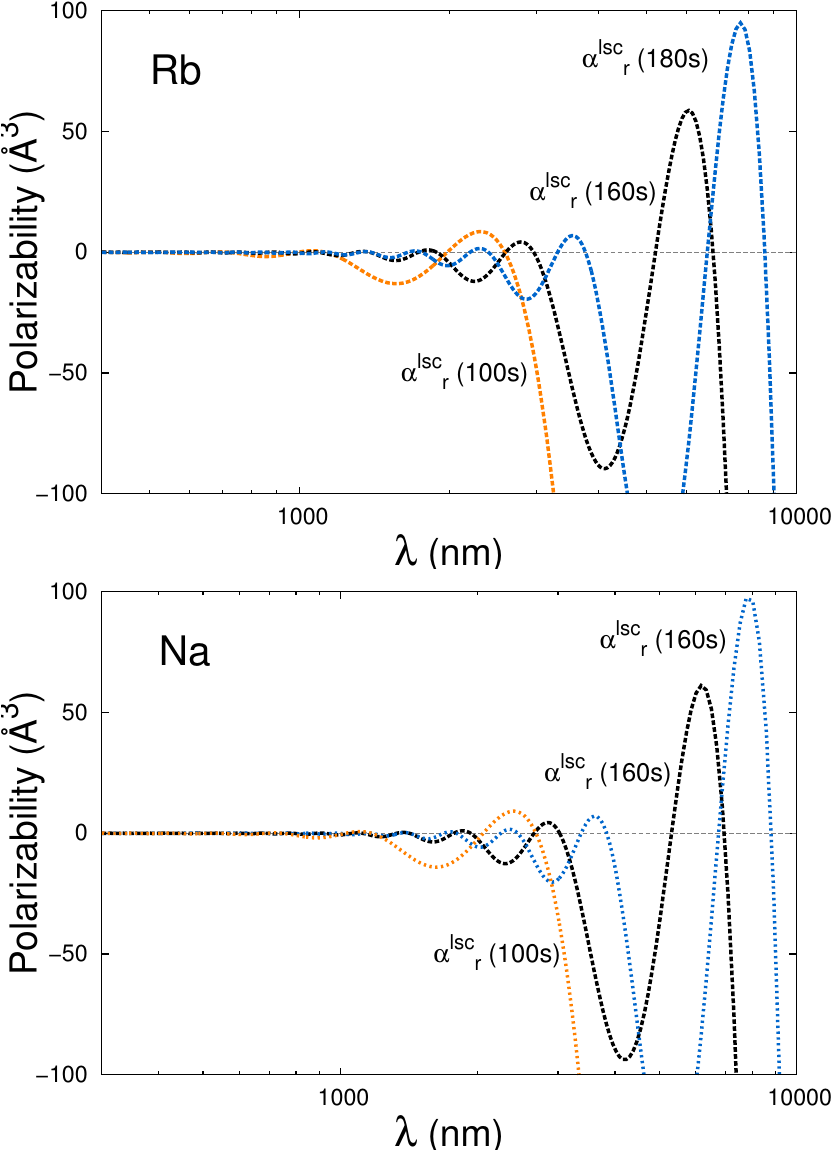}} 
  \end{center}
	\caption{Landscaping polarizabilities $\alpha^{\rm lsc}_{ns}(\omega)$ for 
	$n=100$, 160 and 180 states of the Rb and Na atoms. The $\lambda$ axis is plotted 
	in logarithmic scale to display the range and amplitude of the oscillations. 
	Note that $\sim$10$^4$ nm is the center of the CO$_2$ laser band, and tune-out wavelengths 
	can be found for all $n$ in optical lattices with $\lambda<10^4$ nm. 
	Table~\ref{table:tuneout_wlen} tabulates four largest tune-out wavelengths seen in this 
	figure along with those for the $p$- and the $d$-states. 
	}
	\label{fig:rb_na_tuneout}
\end{figure}

The actual landscaping polarizabilities $\alpha^{\rm lsc}_{n,l,m_z}(\omega)$ calculated 
using~\eqref{cos_expansion} for $n$s-states with $n=100$ (orange), 160 (black) and 180 (blue) 
for the Rb and Na atoms are plotted in Fig.~\ref{fig:rb_na_tuneout}. For all of these states, 
$\alpha^{\rm lsc}_{ns}(\omega)$ oscillates with increasing amplitude before dropping off like 
the free electron polarizability $\alpha_{\rm e}=-1/\omega^2$ before $\lambda\sim 10^4$ nm. 
For all $\lambda <10^4$ nm, there are infinitely many tune-out wavelengths for all Rydberg states 
before the oscillatory nature of $\alpha^{\rm lsc}_{ns}(\omega)$ is dominated by the free 
electron character. Table~\ref{table:tuneout_wlen} tabulates the four largest tune-out wavelengths 
seen in Fig.~\ref{fig:rb_na_tuneout}. Furthermore, it includes the $n=120$s state and the $p$- 
and $d$-states in all of these $n$-manifolds for the Na and Rb atoms. 

The behavior of $\alpha^{\rm lsc}_{n,l,m_z}(\omega)$ seen in Fig.~\ref{fig:rb_na_tuneout} 
can be understood by realizing that $\cos(2kz_e)=1-2\sin^2(kz_e)$, and 
\begin{equation}
\alpha^{\rm lsc}_{n,l,m_z}(\omega) = -\frac{1}{\omega^2} \big( 
	  1 - 2\bra nlm_z|\sin^2(k z_e)|nlm_z \ket \big) . 
\end{equation}
For a given state, $\bra nlm_z|\sin^2(kz_e)|nlm_z\ket\approx 1/2$ at small wavelengths 
and $\alpha^{\rm lsc}_{n,l,m_z}\approx 0$. As the wavelength is increased, $kz_e$ gets smaller 
and $\bra nlm_z|\sin^2(kz_e)|nlm_z\ket\ll 1$ resulting in the free electron polarizability 
dominating $\alpha^{\rm lsc}_{n,l,m_z}$. On the other hand, $kz_e$ becomes larger at a given 
wavelength as $n$ is increased, and $\alpha^{\rm lsc}_{n,l,m_z}$ oscillates until longer 
wavelengths before the free electron behavior takes over. All of these oscillations and 
crossings with the ground state polarizability occur at wavelengths in the CO$_2$ and the doubled 
CO$_2$ laser bands. 

\begin{center}
\begin{figure}[h!tb]
\captionof{table}{Four largest tune-out wavelengths (nm) for the $s$-, $p$- and the $d$-states 
in the $n=100$, 120, 160 and 180 manifolds in Na and Rb atoms. \\}
\begin{tabular}{|C{0.7cm}||C{1.15cm}|C{1.15cm}|C{1.15cm}||C{1.15cm}|C{1.15cm}|C{1.15cm}|}
\hline
 &\multicolumn{3}{c|}{100}&\multicolumn{3}{c|}{120}\\
\cline{2-7}
     & $s$ & $p$ & $d$    & $s$ & $p$ & $d$ \\
\hline\hline
Na   & 2688 \newline 2058 \newline 1171 \newline 1029 
	& 4718 \newline 1825 \newline 1343 \newline 967 
	& 3909 \newline 2629 \newline 1323 \newline 1063 
	& 3888 \newline 2978 \newline 1694 \newline 1488 
	& 6813 \newline 2636 \newline 1939 \newline 1396 
	& 5630 \newline 3787 \newline 1906 \newline 1531  \\ 
\hline
Rb   & 2589 \newline 1983 \newline 1128 \newline 991 
	& 4548 \newline 1759 \newline 1294 \newline 932  
	& 3805 \newline 2559 \newline 1288 \newline 1035 
	& 3768 \newline 2887 \newline 1642 \newline 1443 
	& 6608 \newline 2557 \newline 1881 \newline 1354 
	& 5505 \newline 3702 \newline 1863 \newline 1497  \\ 
\hline
\end{tabular}
\begin{tabular}{|C{0.7cm}||C{1.15cm}|C{1.15cm}|C{1.15cm}||C{1.15cm}|C{1.15cm}|C{1.15cm}|}
\hline
 &\multicolumn{3}{c|}{160}&\multicolumn{3}{c|}{180}\\
\cline{2-7}
     & $s$ & $p$ & $d$    & $s$ & $p$ & $d$ \\
\hline\hline
Na   & 6951 \newline 5324 \newline 3029 \newline 2662 
	& 12156 \newline 4703 \newline 3460 \newline 2492 
	& 10011 \newline 6732 \newline 3388 \newline 2722 
	& 8814 \newline 6752 \newline 3841 \newline 3375 
	& 15403 \newline 5959 \newline 4385 \newline 3157 
	& 12670 \newline 8521 \newline 4288 \newline 3445  \\ 
\hline
Rb   & 6790 \newline 5203 \newline 2959 \newline 2601 
	& 11882 \newline 4597 \newline 3382 \newline 2436 
	& 9843 \newline 6620 \newline 3332 \newline 2677 
	& 8633 \newline 6614 \newline 3762 \newline 3307 
	& 15095 \newline 5840 \newline 4297 \newline 3094 
	& 12482 \newline 8394 \newline 4225 \newline 3394   \\ 
\hline
\end{tabular}
\label{table:tuneout_wlen}
\end{figure}
\end{center}

\section{Three-dimensional optical lattices}

So far we focused on one dimension. 
The position dependent polarizabilities in Fig.~\ref{fig:rb_na_tuneout} are for the $l=0$ 
Rydberg states, and exhibit tune-out frequencies in traps designed for one-dimensional confinement 
along the $z$-direction. If the 1D lattice is formed by counter-propagating Gaussian beams, 
the Rydberg atom may not be trapped in the radial direction. In this case, one could work with 
3D lattices. Since the $s$-electron wave functions are spherically symmetric, the $x$, $y$, and 
$z$ axes are equivalent, and our previous 1D arguments can be directly transferred to the 3D 
lattices. For the $l\neq 0$ states, 3D trapping may not be possible at intensity maxima in all 
directions by proper choice of the lattice wavelength, as the quantization axis and three optical 
beam axes are no longer aligned~\cite{AndRai12}. In this case, the Rydberg atom may be trapped 
at an intensity maximum in one direction, while it may be trapped in intensity minima in other 
directions. 

Formally the 3D lattice problem can be approached in the following fashion: First, we assume 
having six linearly polarized beams in three orthogonal directions. The polarization direction 
(E-field) defines the quantization axis. We then arrange two of the lattices ($x$ and $y$ axes 
would be the optical axes) to have the same polarization along the $z$-axis. If we add the 
third standing wave in the $z$-direction, its polarization will lay in the $x-y$ plane. If we 
choose the frequencies of the optical lattices to be slightly off, the interference effects can 
be made negligible~\cite{GreManTil02}. Then we simply add the optical potentials independently: 
$U^{R}_{n,l,m_z} = U^{X}_{n,l,m_z}+ U^{Y}_{n,l,m_z} + U^{Z}_{n,l,m_z}$, 
\begin{eqnarray}
U^{X}_{n,l,m_z} &=& -\alpha^{\rm lsc}_{n,l,m_z}(\omega;X) \frac{F_0^2}{4} \;, \\
U^{Y}_{n,l,m_z} &=& -\alpha^{\rm lsc}_{n,l,m_z}(\omega;Y) \frac{F_0^2}{4} \;, \\
U^{Z}_{n,l,m_z} &=& -\alpha^{\rm lsc}_{n,l,m_z}(\omega;Z) \frac{F_0^2}{4} \;.
\end{eqnarray} 
Here $X$, $Y$ and $Z$ refer to the lattice coordinates of the atom in the $x$, $y$ and $z$ 
directions. The associated landscaping polarizabilities are 
\begin{eqnarray}
\alpha^{\rm lsc}_{n,l,m_z}(\omega_x;X) &=& -\frac{1}{\omega_x^2} 
		\bra nlm_z|\cos(2k_x x) |nlm_z\ket \;, \label{eq:alp_mz_x} \\
\alpha^{\rm lsc}_{n,l,m_z}(\omega_y;Y) &=& -\frac{1}{\omega_y^2} 
		\bra nlm_z|\cos(2k_y y) |nlm_z\ket \;, \label{eq:alp_mz_y} \\
\alpha^{\rm lsc}_{n,l,m_z}(\omega_z;Z) &=& -\frac{1}{\omega_z^2} 
		\bra nlm_z|\cos(2k_z z) |nlm_z\ket \;. \label{eq:alp_mz_z}
\end{eqnarray}
Because $m_z$ is defined with respect to the $z$-axis, $\cos(2k_x x)$ and $\cos(2k_y y)$ 
will mix the degenerate $m_z$-states in a given $l$-manifold. However, as we discuss 
in Appendix A, this can be circumvented by either a proper choice of the laser frequency 
in the perpendicular directions, or by application of an external magnetic field. 

We will pick the $x$-axis as the polarization axis of the third standing wave. Since the 
$m_z$-substates are defined with respect to the $z$-axis, the atom will be in a linear 
combination of $m_x$ substates in the $x$-direction. To determine the optical potential in the 
$z$-direction, we need to find this linear combination, and then we can apply 
Eq.~\eqref{cos_expansion} since 
$\bra nlm_z|\cos(2k z) |nlm_z\ket \equiv \bra nlm_x|\cos(2k x) |nlm_x\ket$. 
For $s$-states all of this would not matter - {\it i.e.}, $s$-state Rydberg atoms exhibit 
oscillations seen in Fig.~\ref{fig:rb_na_tuneout} and~\ref{fig:cos_scaled} no matter what. Also, 
in general one could pick the quantization axis (defined by external B-field) arbitrarily, and 
see if at some angle the polarizability can be made to vanish. We use the following procedure to 
evaluate the landscaping polarizabilities~\eqref{eq:alp_mz_x} in various ($l,m_z$)-states. 
\begin{eqnarray}
&&L_x ({j,k}) = \bra nlm_{z}| L_{x} |nlm'_{z}  \ket \;, \\
&&\begin{aligned}
L_x ({j,k}) = \frac{1}{2} 
		\big[ 
				&\sqrt{(l-m'_{z})(l+m'_{z}+1)} \delta_{m_{z},m'_{z}+1} \\
				&+ \sqrt{(l-m'_{z}+1)(l+m'_{z})} \delta_{m_{z},m'_{z}-1} 
		\big] \;.
\end{aligned}
\end{eqnarray}
Here we have labeled the states $|nlm_{z}\ket$ and $|nlm'_{z}\ket$ by $j$ and $k$, and used 
\begin{eqnarray}
L_x &=& \frac{1}{2} \left ( L_{+} + L_{-} \right ) \;, \nonumber \\
\bra l'm'| L_{+} |lm \ket &=& \sqrt{(l-m)(l+m+1)} 
		\;\delta_{l,l'} \delta_{m',m+1} \;, \nonumber \\ 
\bra l'm'| L_{-} |lm \ket &=& \sqrt{(l-m+1)(l+m)} 
		\;\delta_{l,l'} \delta_{m',m-1} \;. \nonumber 
\end{eqnarray}
We then diagonalize the matrix $L_{x}$ within the $l$-manifold of interest and obtain the 
eigenvalues and eigenvectors. The matrix $U_{j,\beta}$ whose columns are these eigenvectors is 
our rotation matrix in the Hilbert space: 
\begin{eqnarray}
\alpha^{\rm lsc}_{n,l,m_z}(\omega;X) &=& -\frac{1}{\omega^2} 
		\bra nlm_z|\cos(2k x) |nlm_z\ket \\
	&=& -\frac{1}{\omega^2} \sum_{j} d_{j,j} U^*_{j,\beta} U_{\beta,j} \;,
\end{eqnarray}
where we assumed $\omega_x = \omega_z \equiv \omega$, and $\beta$ indexes the eigenstates 
of $L_x$ within the $l$-manifold. In the end, $\alpha^{\rm lsc}_{n,l,m_z}(\omega;X)$ is 
expressed as a linear combination of $\alpha^{\rm lsc}_{n,l,m_z}(\omega;Z)$ within the same 
$l$-manifold. Here we are exploiting the identity 
$\bra nlm_z|\cos(2k z) |nlm_z\ket \equiv \bra nlm_x|\cos(2k x) |nlm_x\ket$ between the 
matrix elements, which can be evaluated using Eq.~\eqref{cos_expansion}. 

\begin{widetext}
\begin{figure*}[h!tb]
	\begin{center}
 		\resizebox{170mm}{!}{\includegraphics{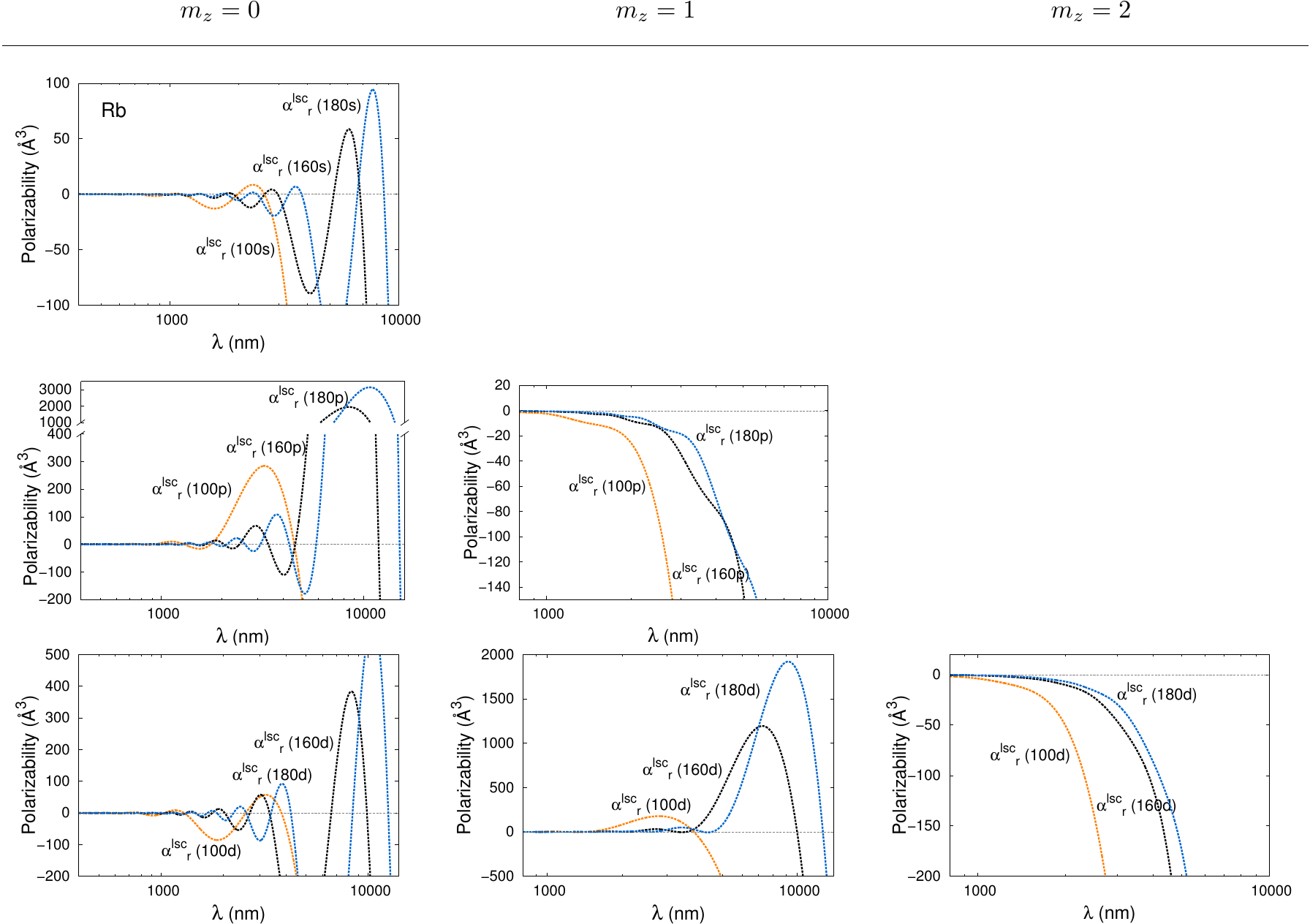}} 
  \end{center}
	\caption{$\alpha^{\rm lsc}_{n,l,m_z}(\omega;Z)$ for $l$ (rows) and $m_z$ (columns) 
	states for the Rb atom. Only $m_z>0$ are shown since $\alpha^{\rm lsc}_{n,l,m_z}(\omega;Z)$ 
	is identical for states with $+m_z$ and $-m_z$. For ($l,m_z$)$=$(0,0), (1,0), (2,0), and (2,1) 
	$\alpha^{\rm lsc}_{n,l,m_z}(\omega;Z)$ oscillates and the angular coefficients 
	in~\eqref{cos_expansion} come in with different signs. For the (1,1) and (2,2) states, 
	it is always negative because the angular coefficients in~\eqref{cos_expansion} are all 
	positive. 
	}
	\label{fig:Rb_3d_z}
\end{figure*}
\end{widetext}

Landscaping polarizabilities $\alpha^{\rm lsc}_{n,l,m_z}(\omega;Z)$ for a one-dimensional 
optical lattice with polarization in the $z$-direction are seen in Fig.~\ref{fig:Rb_3d_z} 
when the Rydberg atom is in $s$-, $p$- and $d$-states. The rows in the figure represent $l=0$, 
1 and 2 from top to bottom, and the columns are the $m_z$ substates. Because 
Eq.~\eqref{cos_expansion} is identical for $\pm|m_z|$ states within the same $l$-manifold, 
we only plot $m_z >0$ states in Fig.~\ref{fig:Rb_3d_z}. For all $m_z=0$ states, the landscaping 
polarizability oscillates and the atom is trapped at intensity minima when 
$\alpha^{\rm lsc}_{n,l,0}(\omega;Z) <0$ and it is trapped at intensity maxima when 
$\alpha^{\rm lsc}_{n,l,0}(\omega;Z) >0$. 
As the wavelength varies, there are wavelengths at which this high and low intensity seeking 
character turns into one another. At these wavelengths, $\alpha^{\rm lsc}_{n,l,0}(\omega;Z)$ 
vanishes and the optical lattice is invisible to the atom. On the other hand, such tune-out 
wavelengths only exist for $p$-states when $m_z =1$, and they are non-existent for the 
extreme $m_z$ states for $l >0$. In these circular states, the landscaping polarizability is 
always negative, and the atom always seeks low intensity and is aware of the trapping potential. 

\begin{widetext}
\begin{figure*}[h!tb]
	\begin{center}
 		\resizebox{170mm}{!}{\includegraphics{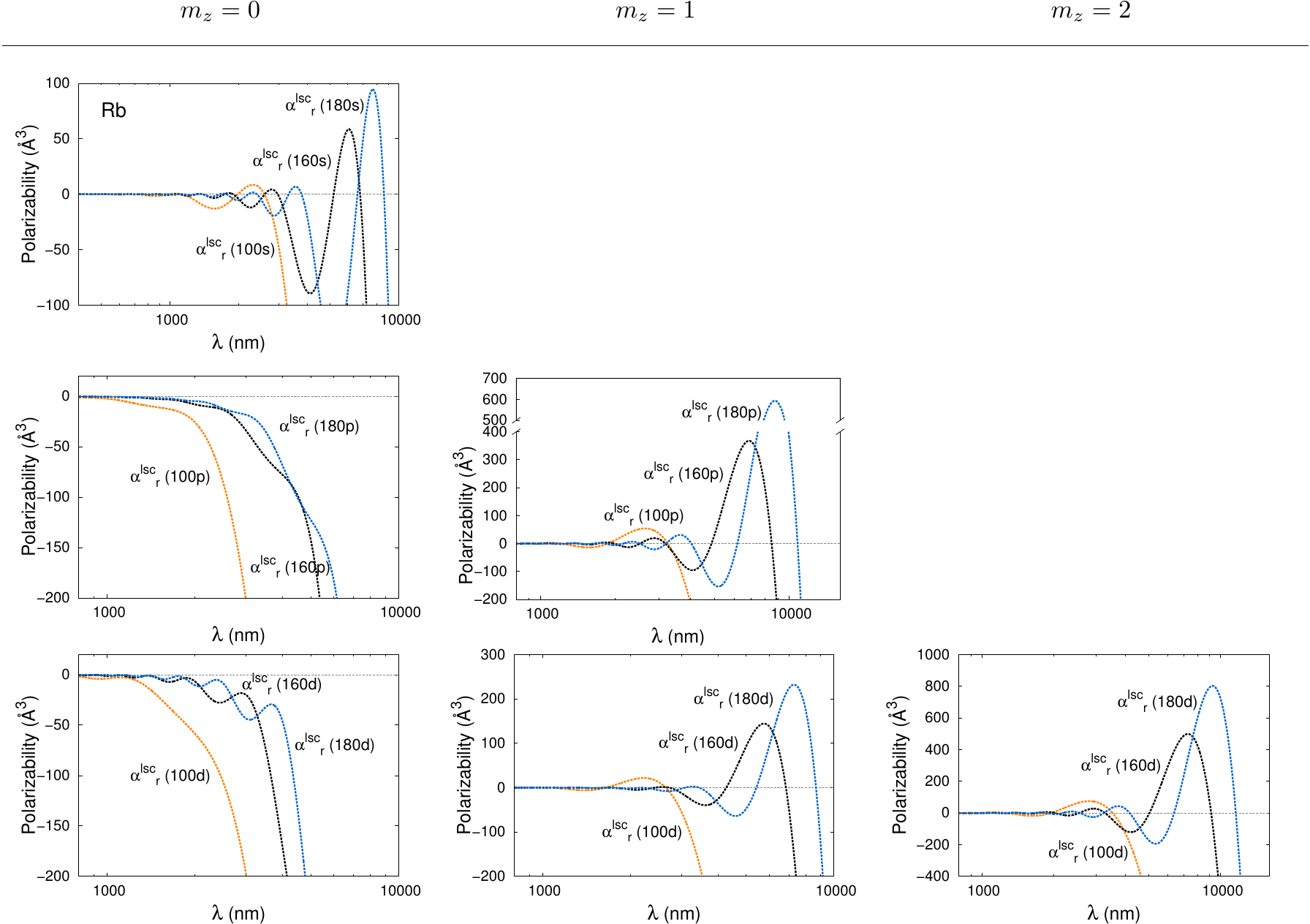}} 
  \end{center}
	\caption{$\alpha^{\rm lsc}_{n,l,m_z}(\omega;X)$ for $l$ (rows), $m_z$ (columns) states 
	for the Rb atom. Only $m_z >0$ are shown because $\alpha^{\rm lsc}_{n,l,m_z}(\omega;X)$ is 
	a linear combination of $\alpha^{\rm lsc}_{n,l,m_z}(\omega;Z)$ within the $l$-manifold 
	when $z$ is chosen to be the quantization axis, and $\alpha^{\rm lsc}_{n,l,m_z}(\omega;Z)$ 
	are identical for states with $+m_z$ and $-m_z$ from Eq.~\eqref{cos_expansion}. Note that 
	$\alpha^{\rm lsc}_{n,l,m_z}(\omega;X)$ and $\alpha^{\rm lsc}_{n,l,m_z}(\omega;Z)$ are 
	the same for $l=0$. 
	}
	\label{fig:Rb_3d_x}
\end{figure*}
\end{widetext}

We now turn on a second lattice with perpendicular polarization to the first one. 
We take the polarization of this second beam to be in the $x$-direction. Then the 
landscaping polarizabilities for the same Rydberg states in Fig.~\ref{fig:Rb_3d_z} 
are seen in Fig.~\ref{fig:Rb_3d_x}. In this case, states with $m_z =\pm|l|$ exhibit 
tune-out conditions for all $l$, and the $m_z=0$ states for $l>0$ behave as the 
circular states in Fig.~\ref{fig:Rb_3d_z}. There are no tune-out wavelengths for these 
states and the atom is always attracted to intensity minima.

\section{Analytical expressions for tune-out wavelengths}\label{sec:analytical}

For individual $l$-states, closed form analytical expressions can be derived for 
Eq.~\eqref{eq:alp_mz_z} using hydrogenic radial wave functions, and the results 
can be generalized for alkali atoms by changing principal quantum number $n$ to 
effective $n^*$ and changing orbital angular momentum $l$ to effective $l^*$ 
(to keep the number of radial nodes fixed). 
We will start with the expression for the landscaping part 
of the Rydberg state polarizability for the case of linearly polarized laser beams where 
$z$-axis is the propagation axis: 
\begin{eqnarray}\label{eq:alp_lsc_exact}
{\alpha}^{\rm lsc}_{n,l,m_z}(k) &=& -\frac{1}{\omega^2} 
		\bra nlm_z|\cos(2kz_{\rm e})|nlm_z\ket \;,
\end{eqnarray}
and $k=2\pi/\lambda$ is the wavevector. The matrix element in~\eqref{eq:alp_lsc_exact} 
can be expressed as an expansion in the spherical Bessel functions $j_{l'}(r_{\rm e})$ as 
in Eq.\eqref{cos_expansion} for given state $|nlm_z\ket$. 
Specifically, for $l=0$, 1 and 2, and $m_z=0$, the landscaping polarizability 
${\alpha}^{\rm lsc}_{n,l,m_z}$ can be expressed as  
\begin{eqnarray*}
{\alpha}^{\rm lsc}_{n,0,0}(k) &=& -\frac{1}{\omega^2} \int_0^\infty dr_{\rm e} 
	P^2_{n,0}(r_{\rm e}) j_0(2kr_{\rm e}) \;, \\
{\alpha}^{\rm lsc}_{n,1,0}(k) &=& -\frac{1}{\omega^2} \int_0^\infty dr_{\rm e} 
	P^2_{n,1}(r_{\rm e}) \big[ j_0(2kr_{\rm e}) - 2j_2(2kr_{\rm e}) \big] \;, \\
{\alpha}^{\rm lsc}_{n,2,0}(k) &=& -\frac{1}{\omega^2} \int_0^\infty dr_{\rm e} 
	P^2_{n,2}(r_{\rm e}) \big[ j_0(2kr_{\rm e}) - \frac{10}{7}j_2(2kr_{\rm e}) \\ 
		&&+ \frac{18}{7}j_4(2kr_{\rm e}) \big] \;.
\end{eqnarray*}
These expressions can be represented with the aid of the definition, 
\begin{eqnarray}\label{eq:def_snlq}
{s}_{n,l,q}(k) \equiv -\frac{1}{\omega^2} \int_0^\infty dr_{\rm e} 
	P^2_{n,l}(r_{\rm e}) j_q(2kr_{\rm e}) \;,
\end{eqnarray}
so that the above expressions for the landscaping polarizabilities read : 
\begin{eqnarray}
{\alpha}^{\rm lsc}_{n,0,0}(k) &=& {s}_{n,0,0}(k)\label{s_state} \;, \\
{\alpha}^{\rm lsc}_{n,1,0}(k) &=& {s}_{n,1,0}(k) - 2{s}_{n,1,2}(k) \;, \\
{\alpha}^{\rm lsc}_{n,2,0}(k) &=& {s}_{n,2,0}(k) -\frac{10}{7}{s}_{n,2,2}(k) 
	+ \frac{18}{7} {s}_{n,2,4}(k) \;.
\end{eqnarray}
We derive closed form expressions for ${s}_{n,l,q}$ in Appendix B using 
the well known radial wave functions for the hydrogen atom. 

\subsection{Landscaping polarizabilities and tune-out wavelengths for $s$-states}
Since $l=0$ is the simplest case (Eq.~\eqref{s_state}), 
we would like to obtain an approximate expression for ${\alpha}^{\rm lsc}_{n,0}$ and 
the tune-out wavelengths at which it vanishes. 
For an $s$-state, ${\alpha}^{\rm lsc}_{n,0}={s}_{n,0,0}$ and therefore $q=0$. 
In our derivation, we will only keep the leading term 
in the sum~\eqref{alpha_sum}. This will undermine the oscillatory character of the Rydberg state 
and we find that it yields an accurate expression for half of the tune-out frequencies.  The 
reason that ignoring the oscillations of the Rydberg orbital yields us anything acceptable is that 
the main contribution to the integral in~\eqref{cos_expansion} comes from near the classical 
turning point, where the oscillatory parabolic nature of the Rydberg orbital assumes an exponentially 
decaying hyperbolic form. By only taking the largest order term in the Laguerre polynomials, we are 
still ensuring that this part of the wavefunction is included in our approximate Rydberg orbital, 
albeit with overestimated amplitude. 
Thus, keeping the leading term in the sum~\eqref{alpha_sum}, we obtain 
\begin{eqnarray}\label{leading} 
{\alpha}^{lsc}_{n,0} \simeq -A_{n^*l^*} \frac{(1+\xi^2)^{-n^*}}{\xi^3} 
	\sin\big[2n^*\tan^{-1}\xi \big] \;,
\end{eqnarray} 
where $A_{n^*l^*}$ is a coefficient. This expression vanishes when the argument of the 
sine function is an integer multiple of $\pi$, meaning that the 
atom will not feel the optical lattice when 
\begin{eqnarray}\nonumber
2n^*\tan^{-1}\xi^{\otimes} &=& p \pi\;\;\;(p=1,2,\cdots) \\
\xi^{\otimes} &=& k^{\otimes} n^* = \left | \tan\Big(\frac{p \pi}{2n^*} \Big) \right |. 
\end{eqnarray}
The index $p$ counting the roots start from 1 because $p=0$ results in infinitely long 
wavelength. In increasing order, $p$ yields half of the available tune-out wavelengths 
from longer to shorter wavelengths. Thereby the tune-out wavelengths are 
\begin{equation}\label{eq:roots}
\lambda^{\otimes}_p = \frac{2\pi n^{*}}{\tan [p \pi/(2n^*)]} \;. 
\end{equation} 
For large $n^{*}$ (or small values of $p \pi/(2n^*)$), this has the scaling 
$\lambda^{\otimes}_p \sim 4(n^{*})^2 /p$. 

\begin{center}
\begin{figure}[h!tb]
\captionof{table}{Calculated $\lambda^{\otimes}$ versus the analytically estimated 
$\lambda^{\otimes}_{p}$ (nm) from Eq.~\eqref{eq:roots} for 160s and 180s states of Na and 
Rb. $n=160$ refers to $n^* = 158.58$ and 156.76, and $n=180$ refers to $n^* = 178.58$ and 
176.75 for Na and Rb, respectively. \\} 
\begin{tabular}{|C{0.7cm}||C{1.5cm}|C{1.5cm}||C{1.5cm}|C{1.5cm}|} 
\hline
 &\multicolumn{2}{c|}{160}&\multicolumn{2}{c|}{180}\\
\cline{2-5}
     & $\lambda^{\otimes}$ & $\lambda^{\otimes}_p$    \
     & $\lambda^{\otimes}$ & $\lambda^{\otimes}_p$ \\ 
\hline\hline
Na   & \begin{tabular}{c} 
					6951 \\ \underline{5324} \\ 3029 \\ \underline{2662} 
			 \end{tabular}
	& \begin{tabular}{c} 
				  \\ \underline{5323} \\  \\ \underline{2661}
		\end{tabular} 
	& \begin{tabular}{c}  
				8814 \\ \underline{6752} \\ 3841 \\ \underline{3375} 
		\end{tabular}
	& \begin{tabular}{c} 
				\\ \underline{6750} \\ \\ \underline{3375} 
		\end{tabular} \\ 
\hline
Rb  & \begin{tabular}{c}  
					6790 \\ \underline{5203} \\ 2959 \\ \underline{2601} 
			\end{tabular}
	& \begin{tabular}{c}  
				\\ \underline{5201} \\ \\ \underline{2600}
		\end{tabular}
	& \begin{tabular}{c}  
				8633 \\ \underline{6614} \\ 3762 \\ \underline{3307} 
		\end{tabular}
	& \begin{tabular}{c}  
				\\ \underline{6613} \\ \\ \underline{3306} 
		\end{tabular} \\ 
\hline
\end{tabular}\label{table:analytic_wlen}
\end{figure}
\end{center}

In Table~\ref{table:analytic_wlen}, we present a comparison of the analytical and the numerically 
evaluated $\lambda^{\otimes}$. The tune-out wavelengths predicted by~\eqref{eq:roots} are 
tabulated in Table~\ref{table:analytic_wlen} together with the exact values for 160s and 180s 
states of Na and Rb. Note that our analytical expression only predicts every other available tune-out 
wavelength, although it is a very good approximation for the ones it can predict. The agreement 
for these $\lambda^{\otimes}$ is better than a part in a thousand for both Na and Rb atoms. 
Although taking only the leading term in the Laguerre polynomial to approximate the radial part 
of the integral in~\eqref{cos_expansion} is a very crude approximation (because it ignores the 
oscillatory nature of the Rydberg wavefunction), this approximation predicts half of the tune-out 
wavelengths surprisingly well. 

\section{Conclusions}
Many recent advances in precision time keeping, quantum control with cold molecules and quantum 
computation science rely on trapping cold atoms and molecules in optical lattices. The common wisdom 
is that alkali atoms in Rydberg states see trapping potentials that are essentially that of a free 
electron, and are always low intensity seekers in an optical lattice. The change in the high and low  
intensity seeking character of the Rydberg states results from a non-monotonic interplay between the 
lattice constant and the physical size of the Rydberg state. Recently, we demonstrated that this 
view needs revision, and Rydberg atoms can be made to be both low and high intensity seekers by 
proper choice of lattice wavelength~\cite{TopDerLands}. In other words, the lattice potential can 
be positive or negative for a given Rydberg state. To change sign, the lattice potential has to 
vanish at certain wavelengths at which the optical lattice becomes invisible to the atom. In this 
paper, we have illustrated this phenomenon through extensive calculations. At these frequencies, 
the atom in the targeted state is ``tuned out" of the lattice, while atoms in other Rydberg states 
are still trapped at either intensity minima or maxima. We have provided a list of these tune-out 
wavelengths for a few highly excited states of alkali metal atoms Na and Rb in one-dimensional IR 
lattices. 

We then turn our attention to optical lattices beyond one-dimension, and demonstrate that the 
optical lattice potential does not change sign for all Rydberg states $|nlm_z \ket$ for a given 
lattice polarization by simply tuning the lattice wavelength. For example, for a one-dimensional 
lattice formed by two laser beams propagating along the $z$-axis with polarization in the 
$x$-direction, the landscaping polarizability exhibits tune out wavelengths only for the $m_z=0$ 
states and the $m_z=\pm1$ states for $l=2$. The optical potential does not change sign for 
any of the circular states with $l>1$. In contrast, when we turn on a second standing wave 
propagating along the $x$-axis with polarization in the $z$-direction, only the $l=0$ and the 
extreme $|m_z|$ states exhibit tune-out wavelengths. In this two-dimensional setup, tune-out 
wavelengths exist simultaneously in both lattices only for the 
$s$-states and $d$-states with $m_z=\pm1$. 

We also derive closed form analytic expressions for the tune-out wavelengths of alkali metal atoms 
in $m_z=0$ Rydberg states trapped in IR optical lattices. Using these expressions, we obtain a 
simple formula for $s$-states, which accurately predicts half of the available tune-out wavelengths. 

\section{Acknowledgments}
This work was supported by the National Science Foundation (NSF) Grant No. PHY-1212482 
and PHY05-25915. 

\section{Appendix A}

Here we demonstrate that in the presence of a second standing wave with polarization 
perpendicular to the $z$-direction, the mixing between different degenerate $m_z$ sublevels 
in a Ry $l$-manifold can be prevented by a proper choice of the laser frequency. This is 
particularly useful when there are no tune-out conditions in the second direction, 
and the lattice wavelength can be chosen freely without the tune-out constraint. This 
situation occurs in Fig.~\ref{fig:Rb_3d_x} for several $m_z$-states. When the tune-out 
condition also needs to be achieved in the second direction, an external magnetic field 
can be applied to split the $m_z$ sublevels so that the splitting is much larger than the 
matrix elements connecting different $m_z$ states. This would effectively prevent $m_z$ 
mixing inside a Ry $l$-manifold. 
In this section, we will 
assume that the second standing wave is polarized in the $x$-direction. 

Because $\sin(2kx_e)$ mixes different $m_z$ substates, it is necessary to construct and 
diagonalize the matrix 
\begin{eqnarray} 
\mathbf{U}^{X}_{r} = -\frac{F_0^2}{4} \pmb{\alpha}(\omega;X) \;,
\end{eqnarray}
to obtain the optical potential in the second direction. Here $\pmb{\alpha}(\omega;X)$ is 
the landscaping polarizability matrix, and the matrix elements of $\mathbf{U}^{X}_{r}$ and 
$\pmb{\alpha}(\omega;X)$ are 
\begin{eqnarray} 
&&U^{X}_{r}(nlm_z \;;\; n'l'm_z') = -\frac{F_0^2}{4} \alpha^{n'l'm'}_{n,l,m}(\omega;X) \;, \\ 
&&\alpha^{n'l'm'}_{n,l,m}(\omega;X) \equiv -\frac{1}{\omega^2} 
		\bra nlm|\cos(2kx_e)|n'l'm'\ket \;.
\end{eqnarray}
For example, $\mathbf{U}^X_r$ is a $3\times 3$ matrix for a $p$-state where only the 
diagonal and $U^{X}_{r}(n,1,\pm 1 ; n,1,\mp 1)$ off-diagonal matrix elements 
are non-zero. 

The operator we have been using to evaluate the landscaping polarizability in the $z$-direction is 
\begin{eqnarray}\label{appA_cos2kz} \nonumber 
\cos(2kz_e) &=& \sum_{L=even} (2L+1)(-1)^{L/2} j_L(2kr_e) P_L(\cos\theta) \\ 
	&=& \sum_{L=even} a_{L,0}(r) Y_{L,0}(\hat{r}) \;,
\end{eqnarray}
where $a_{L,0}(r) = (-1)^{L/2} \sqrt{4\pi(2L+1)} j_L(2kr_e)$, $j_L$ are spherical Bessel 
functions, and $Y_{L,M}$ are the spherical harmonics. We would like to rotate $z_e$ into $x_e$ 
such that $\cos(2kz_e)\rightarrow \cos(2kx_e)$, where $\widehat{S}$ is an operator 
that rotates $z_e$ to $x_e$. The effect of rotation on the spherical harmonics can be expressed 
in terms of the Wigner D-functions~\cite{VarMosKhe89}: 
\begin{eqnarray}\label{appA_ylm} 
Y_{L,M'}(\hat{S}\hat{r}) &=& \sum_{M=-L}^{L} Y_{L,M}(\hat{r}) 
	D_{MM'}^{L}(\alpha,\beta,\gamma) \;,
\end{eqnarray}
where $(\alpha,\beta,\gamma)$ are the Euler angles. With the choice of 
angles $(0,\pi/2,0)$, the operator $\cos(2kx_e)$ can be written as 
\begin{eqnarray} \nonumber
\cos(2kx_e) &=& \sum_{L,M} a_{L,0}(r) Y_{L,M}(\hat{S}\hat{r}) \\ \nonumber
	&=& \sum_{L,M} a_{L,0}(r)  Y_{L,M}(\hat{r}) 
		D_{M0}^{L}(0,\pi/2,0) \;,
\end{eqnarray}
where $L$ is even. From Ref.~\cite{VarMosKhe89} 
\begin{eqnarray}\label{appA_Dm0} 
D_{M0}^{L}(\alpha,\beta,\gamma) &=& \sqrt{\frac{4\pi}{2L+1}} Y^*_{LM}(\beta,\alpha) \;,
\end{eqnarray}
which leads to 
\begin{eqnarray} 
\cos(2kx_e) = \sum_{L,M} T^{(L)}_{M} 
		\sqrt{\frac{4\pi}{2L+1}} Y^*_{LM}(\pi/2,0) \;.
\end{eqnarray}
Here we have defined $T^{(L)}_{M}\equiv a_{L,0}(r)Y_{L,M}(\hat{r})$. The 
matrix elements of $\pmb{\alpha}(\omega;X)$ therefore become 
\begin{eqnarray} 
\alpha^{n'l'm'}_{n,l,m}(\omega;X) \equiv -\frac{1}{\omega^2} \sum_{L,M} 
		B_{L,M} \bra nlm|T^{(L)}_M|n'l'm'\ket \;.
\end{eqnarray}
For the specific Euler angles $\beta=\pi/2$ and $\alpha=0$, a closed form expression 
can be found for $Y^*_{LM}(\pi/2,0)$~\cite{VarMosKhe89}, and the coefficients $B_{L,M}$ 
become 
\begin{eqnarray} 
\resizebox{.99\hsize}{!}{$
B_{L,M} = \begin{cases} 
      (-1)^{\tfrac{L+M}{2}} \sqrt{\tfrac{(L+M-1)!!}{(L+M)!!} 
      	\tfrac{(L-M-1)!!}{(L-M)!!} } & L+M \;\;\;\rm{even}  \\ \nonumber
      0 & L+M \;\;\;\rm{odd} 
   \end{cases}$ } \;.
\end{eqnarray}
Using the Wigner-Eckart theorem, we can express the matrix elements for the operator 
$T^{(L)}_M$ in terms of the reduced matrix elements: 
\begin{align} \nonumber
\bra nlm_z |T^{(L)}_M &|n'l'm_z'\ket = (-1)^{l-m_z} \\ 
		&\times \threej{l}{L}{l'}{-m_z}{M}{m_z'} 
		\bra nl ||T^{(L)}||n'l'\ket \;,
\end{align}
where 
\begin{align} \nonumber
\bra nl ||T^{(L)}&||n'l'\ket = (2L+1) (-1)^{\tfrac{L}{2} -l} 
	\sqrt{(2l+1)(2l'+1)} \\ \nonumber
	&\times \threej{l}{L}{l'}{0}{0}{0} 
		\int_0^\infty dr_{e} P_{nl}(r_e) j_{L}(2kr_e) P_{n'l'}(r_e)\;.
\end{align}
%
%
This method of evaluating the landscaping polarizabilities 
$\alpha^{\rm lsc}_{n,l,m}(\omega;X)$ reproduces Fig.~\ref{fig:Rb_3d_x}, which 
was calculated by rotating the eigenstates. The diagonal matrix elements of 
$\pmb{\alpha}(\omega;X)$ correspond to $\alpha^{\rm lsc}_{n,l,m}(\omega;X)$. 

\begin{figure}[h!tb]
	\begin{center}
 		\resizebox{85mm}{!}{\includegraphics{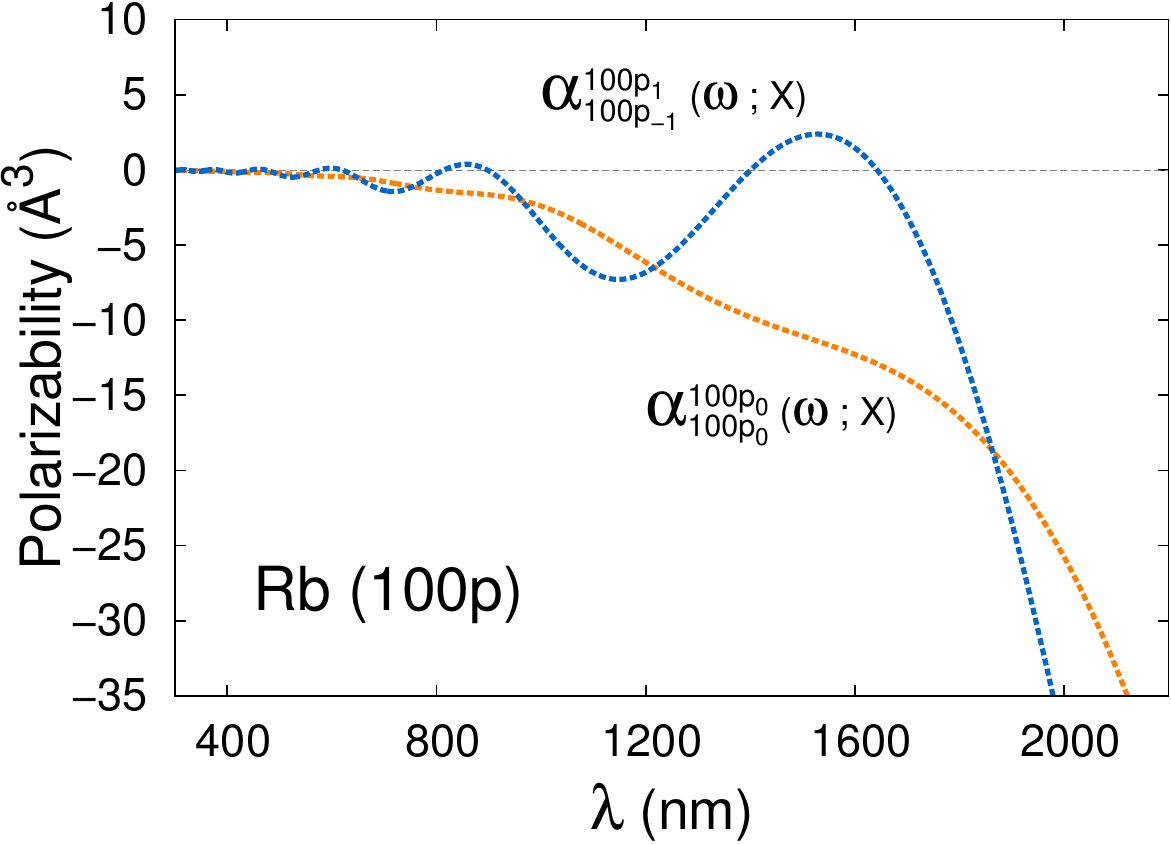}} 
  \end{center}
	\caption{(Color online) Diagonal $\alpha^{100p_0}_{100p_0}(\omega;X)$ (dashed orange) 
	and off-diagonal $\alpha^{100p_1}_{100p_{-1}}(\omega;X)$ (dashed blue) matrix elements 
	of the landscaping polarizability matrix $\pmb{\alpha}(\omega;X)$. 
	The off-diagonal matrix element vanishes at several wavelengths. 
	}
	\label{fig:rb_appendix}
\end{figure}

Evaluating the diagonal and off-diagonal matrix elements of the landscaping polarizability 
matrix $\pmb{\alpha}(\omega;X)$, we see that the off-diagonal matrix 
elements can be comparable to the diagonal elements at certain wavelengths. On the other 
hand, the off-diagonal matrix elements can be also  made to vanish at several wavelengths. 
For example, in Fig.~\ref{fig:rb_appendix} we plot 
the $m_z =0$ diagonal matrix element and the only non-zero off-diagonal element for the 
100p state of Rb. 
This diagonal matrix elements is identical to the landscaping polarizability 
seen in the second row of Fig.~\ref{fig:Rb_3d_x} for $m_z =0$. It is also clear that the 
diagonal matrix element vanishes at several wavelengths, such as 1642, 1401, 897, and 815 nm. 
At these wavelengths $\pmb{\alpha}(\omega;X)$ is diagonal, therefore there is no mixing 
between the different $m_z$-states of the 100p manifold. 

Alternatively, one can apply an external magnetic field $B$ to lift the degeneracy between 
$m_z$ states such that the Zeeman splitting is much larger than the off-diagonal matrix 
elements. In this case, the splitting occurs only between the diagonal matrix elements, and 
the optical potential becomes 
$\mathbf{U}^{X}_{r} + m\mu_{\rm B} B \;\mathbf{I}$, where $\mu_B$ is the 
magnetic moment and $\mathbf{I}$ is the identity matrix. This means that when magnetic 
field is chosen so that $\mu_{\rm B} B \gg U^{X}_{r}(n,1,\pm 1 ; n,1,\mp 1)$, where 
$U^{X}_{r}(n,1,\pm 1 ; n,1,\mp 1)$ is comparable to the diagonal matrix elements, the optical 
potential can be made essentially diagonal, effectively preventing mixing between the $m_z$ 
sublevels of the $l$-manifold. For example, for a Ry state in a 10 mK deep optical 
trap, the external magnetic field needed to split the degenerate levels by 10 mK is $\sim150$ 
Gauss. Similar considerations apply for an optical lattice oriented in the $y$-direction.

\section{Appendix B}
In order to derive closed form analytical expressions for $s_{n,l,q}$ defined in 
Sec.~\ref{sec:analytical}, we use the well known radial wave functions for the hydrogen atom 
in terms of the associated Laguerre polynomials~\cite{gallagher}: 
\begin{align} \nonumber
R_{n,l}(r_{\rm e}) = \frac{2}{n^2} &\sqrt{\frac{(n-l-1)!}{(n+l)!}} \\
	&\left(\frac{2r_{\rm e}}{n}\right)^l \nonumber 
		e^{-r_{\rm e}/n} L_{n-l-1}^{2l+1} \left(\frac{2r_{\rm e}}{n}\right) \;.
\end{align}
Then by substituting this into Eq.~\eqref{eq:def_snlq} and defining $\xi=kn$, $\rho=2r_{\rm e}/n$ 
and simplifying, we get
\begin{eqnarray}\label{hyd_wf} 
{s}_{nl}(\xi) &=& -\frac{n^2}{c^2 \xi^2} 
		\frac{\left(n-l-1\right)!}{2n\left(n+l\right)!} \\ \nonumber
		&&\times \int_0^\infty e^\rho \rho^{2(l+1)} [L_{n-l-1}^{2l+1}(\rho)]^2 j_q(\xi\rho) 
			d\rho \;, 
\end{eqnarray}
where $[L_{n-l-1}^{2l+1}(\rho)]^2$ is the square of the Laguerre polynomial~\cite{bailey}: 
\begin{widetext}
\begin{eqnarray}\label{slaguerre}
[L_{n_r}^\alpha(\rho)]^2 = \frac{({n_r}+\alpha)!}{2^{2{n_r}}{n_r}!}\sum_{j=0}^{{n_r}} 
	\frac{(2j)!\:(2{n_r}-2j)!}{j!\:\{({n_r}-j)!\}^2\:(\alpha+j)!} 
		\sum_{k=0}^{2j} \frac{(2j+2\alpha)!\:2^k}{(2j-k)!\:(2\alpha+k)!} 
		\frac{(-1)^k}{k!} \rho^k 
\end{eqnarray}
\end{widetext}
with $n_r=n-l-1$ (number of radial nodes) and $\alpha=2l+1$. Eqs.~\eqref{hyd_wf} 
and~\eqref{slaguerre} can be used to cast 
${s}_{nl}(\xi)$ into a finite sum of the form 
\begin{eqnarray}\label{alpha_sum}
{s}_{n,l,q}(\xi) = \frac{1}{k^3} \sum_{p=0}^{2n_r} b_p \: I_p(q) \;, 
\end{eqnarray}
where $I_p(q)$ involves the integrals over $\rho$, and $b_p$ are coefficients 
involving quantum numbers. We now derive coefficients $b_p$ and the integrals $I_p(q)$. 

Expression~\eqref{slaguerre} 
is a nested double sum, where the power of $\rho$ is $k$, the index of the {\it inner} sum. 
Therefore, Eq.~\eqref{slaguerre} gives a polynomial in a form which has more than one term 
involving a given power of $\rho$. To obtain the coefficients $b_j$, we need an expression 
in which the powers of $\rho$ are in terms of the index of the {\it outer} summation. 
This can be done by rearranging the terms in Eq.~\eqref{slaguerre} in the following fashion: 
\begin{widetext}
\begin{eqnarray}\label{rearranged}
&&\begin{aligned} 
	[L_{n_r}^\alpha(\rho)]^2 = 
		\frac{({n_r}+\alpha)!}{2^{2{n_r}}{n_r}!} \sum_{p=0}^{2{n_r}} \sum_{g=0}^{G(p)} 
		&\frac{(2{n_r}-2g)!\:(2g)!}{({n_r}-g)!\:\{g!\}^2\:(\alpha+{n_r}-g)!} \\
		&\;\;\;\times \frac{(2{n_r}-2g+2\alpha)!\:2^{2{n_r}-p}}{(p-2g)!\:(2\alpha+2{n_r}-p)!}\frac{(-1)^{2{n_r}-p}}{(2{n_r}-p)!}
		\rho^{2{n_r}-p} \;,
\end{aligned}\\
&&G(p) = \frac{1}{2}\Big[p-\tfrac{1}{2}(1-(-1)^p)\Big ] \;. 
\end{eqnarray}
\end{widetext}
Notice that $G(p)$ is the upper limit of the inner sum in Eq.~\eqref{rearranged}. 
Although these expressions are for the hydrogen atom, they can be generalized to the case 
of an alkali atom utilizing the quantum defect theory such that the energy of a given state 
in an alkali atom fits the hydrogenic form $-1/[2(n^*)^2]$~\cite{gallagher}, and the number 
of radial nodes $n_r=n-l-1=n^*-l^*-1$ is fixed~\cite{kostelecky}. This produces a new set of 
quantum numbers ($n^*$ and $l^*$) for the alkali atom, and since they are no longer 
integers, the factorials in~\eqref{rearranged} must be expressed in terms of the Gamma 
function when necessary, {\it i.e.}, $\Gamma(x)=(x-1)!$. Following through this procedure 
and using the rearranged version of the squared Laguerre polynomials given by 
Eq.~\eqref{rearranged}, ${s}_{n,l,q}$ can be written as 
\begin{widetext}
\begin{eqnarray}
&&\begin{aligned}
{s}_{n,l,q}(\xi) = \frac{n^*}{2c^2 \xi^3} \frac{1}{2^{2n_r}} 
	\sum_{p=0}^{2n_r} 
	\Bigg(
	\sum_{g=0}^{G(p)} 
	&\frac{\Gamma(2n_r-2g+1)\:(2g)!}{\Gamma(n_r-g+1)\:\{g!\}^2\:\Gamma(\alpha^*+n_r-g+1)} \\ \nonumber 
	&\;\;\times \frac{\Gamma(2n_r-2g+2\alpha^*+1)\:2^{2n_r-p}}{(p-2g)!\:\Gamma(2\alpha^*+2n_r-p+1)} 
		\frac{(-1)^{2n_r-p}}{\Gamma(2n_r-p+1)} 
	\Bigg ) \\ \nonumber 
	&\;\;\times \underbrace{\int_0^\infty e^{-\rho} \rho^{2n^*-p} j_q(\xi\rho) d\rho}_{I_p(q)} \;.
\end{aligned} 
\end{eqnarray}
On identifying with Eq.~\eqref{alpha_sum}, we find 
\begin{eqnarray*}
&&n_r = n-l-1 = n^* -l^* -1 \:\:\:\text{and}\:\:\:  \alpha^* = 2l^* +1 \;, \\
&&\begin{aligned}
b_p = \frac{c}{(n^*)^2 2^{2n_r +1}} \sum_{g=0}^{G(p)} 
	&\frac{\Gamma(2n_r-2g+1)\:(2g)!}{\Gamma(n_r-g+1)\:\{g!\}^2\:\Gamma(\alpha^*+n_r-g+1)} \\ \nonumber
	&\;\;\times \frac{\Gamma(2n_r-2g+2\alpha^*+1)\:2^{2n_r-p}}{(p-2g)!\:\Gamma(2\alpha^*+2n_r-p+1)}
	\frac{(-1)^{2n_r-p}}{\Gamma(2n_r-p+1)} \;,
\end{aligned}\\
&&I_p(q) = \frac{\sqrt{\pi}}{2^{q+1}}\: \xi ^q \,\Gamma (2 n-p+q+1) 
	_2\tilde{F}_1\left(\frac{1}{2} (2n-p+q+1),\frac{1}{2} 
	(2n-p+q+2);q+\frac{3}{2};-\xi ^2\right) \;,
\end{eqnarray*}
\end{widetext}
where $_2 \tilde{F}_1$ is the regularized hypergeometric function, and $\xi=kn$.



\end{document}